\documentclass[11pt]{article}
\usepackage[left=3.17cm,top=2.54cm,right=3.17cm,bottom=2.54cm,nohead,nofoot]{geometry}
\usepackage{setspace}
\usepackage{amsmath}
\usepackage{amssymb}
\usepackage{graphicx}
\date{}
\begin{document}
\title{EPR Secure Non-relativistic Bit Commitment Through Entanglement Breaking Channels}
\author{S. Arash Sheikholeslam
sasheikh@ece.uvic.ca\\
Department of Electrical and Computer Engineering, University of Victoria\\
PO Box 3055, STN CSC, Victoria, BC Canada V8W 3P6\\
\and  T. Aaron Gulliver
 \\agullive@ece.uvic.ca
\\
Department of Electrical and Computer Engineering, University of Victoria,\\
PO Box 3055, STN CSC, Victoria, BC Canada V8W 3P6}
\maketitle

\begin{abstract}
This paper considers the use of
an entanglement breaking channel in the construction of a secure bit-commitment protocol.
It is shown that this can be done via a depolarizing quantum channel.
\end{abstract}

\pagebreak

\section*{Introduction}

The most successful cheating strategy against non-relativistic bit commitment schemes is the entanglement attack
(also known as the EPR attack) \cite{Mayers99thetrouble} \cite{PhysRevLett.78.3414}.
In this strategy, one of the parties (Alice)
entangles a system with the one she uses for commitment and keeps this second system secret.
Then she is able to cheat before the opening phase through local operations on her own system.
One approach to counter this cheating strategy is to determine a means of breaking the entanglements.
This must be done either through local transforms performed by the other party (Bob), or through local noise applied to Bob's
system (from the transmission channel).

Entanglement breaking channels are a relatively new concept in quantum information that were first introduced in \cite{15}.
The characteristics of two-qubit entanglements are discussed in \cite{16} and \cite{17}.
In particular,
the local two-qubit entanglement-annihilating channel (2-LEA) is examined in \cite{16}.
>From \cite{15}, a local channel $c$ is called entanglement breaking if the output of the channel operating on an
entangled state is separable, where separability for a density matrix $\rho$ means $\rho=\sum_{i} p_i \rho_a^i\otimes \rho_b^i$.
In the next section, we describe through an example how an entanglement breaking channel can be used to secure the Bennett and
Brassard bit commitment scheme \cite{1984-175-179} against an EPR attack.

\section*{Depolarizing Channel Bit Commitment}

As is typical, we assume Alice is working in a noise free environment, i.e., a perfectly shielded and isolated lab.
Therefore the joint noise which corrupts the entangled state $\rho_{AB}$ is $I \otimes \varepsilon_{c}[\rho_{AB}]$,
where $\varepsilon_{c}$ is the channel noise.
This entanglement breaking operation must either be applied by Bob through some apparatus
he possesses for adding noise, or by the quantum channel through which Alice sends the qubits to Bob, as shown in Figure 1.
%\begin{figure}
%\centering\includegraphics[scale=0.5]{fig1.jpeg}
%\caption{Two possible implementations of the depolarizing channel for bit commitment.}
%\end{figure}
Here we use a depolarizing channel to apply the entanglement breaking operation.
This channel is defined in \cite{17} as
\[
\epsilon(X)=qX+(1-q)tr[X]\frac{1}{2}I
\]
The action of the depolarizing channel replaces the qubit with the completely mixed state, $\frac{I}{2}$, with probability $1-q$.

It was shown in \cite{18}\cite{19} that the evolution of any entangled state in a channel (entanglement breaking channel in this case),
is determined by the evolution of a maximally entangled state in the channel.
Therefore we only consider the effect of the entanglement breaking channel on the maximally
entangled state $\vert \psi^+\rangle=\frac{1}{\sqrt{2}}(\vert0_A0_B\rangle+\vert1_A1_B\rangle)$,
where the subscripts $A$ and $B$ denote Alice and Bob, respectively.
It was proven in \cite{19} for a local quantum channel $\mathbb{S}$ (which operates on a qubit), an entangled state $\vert X\rangle$
(which is a bipartite $N \otimes 2$ state), and concurrence $C$ as defined in \cite{19} (a measure of entanglement), that
\[
C((I\otimes \mathbb{S})[\vert X\rangle\langle X\vert])=C[\vert X\rangle\langle X\vert]]C[(I\otimes \mathbb{S})[\vert \psi^+\rangle \langle \psi^+ \vert]].
\]
Since $C[\vert X\rangle\langle X\vert]]=0$ for $\vert X\rangle$, which is a separable state,
if a local channel $\mathbb{S}$ (or $I\otimes \mathbb{S}$ for the entire system) applied on the
maximally entangled state $\vert \psi^+\rangle$ disentangles it, then we have $C((I\otimes \mathbb{S})[\vert X\rangle\langle X\vert])=0$,
which disentangles any bipartite $N \otimes 2$ state.

Having a maximally entangled state $\vert \psi^+\rangle$ and applying a depolarizing channel on Bob's state (which means the effect of the channel on the
entire system is $(I\otimes\epsilon)[X]$), results in the state
\[
q \vert \psi^+\rangle \langle \psi^+ \vert+\frac{(1-q)}{4}I_A\otimes I_B
\]
which has been shown to be separable for $q \leq \frac{1}{3}$ \cite{16}.
Therefore as discussed above, this channel will disentangle Bob's qubit from any other system (such as Alice's secret system).
Now assume the two parties use the simple Bennett and Brassard bit commitment scheme in which Alice sends random selections of $\vert \uparrow \rangle$
or $\vert \rightarrow\rangle$ for 0, and $\vert \nearrow \rangle$ or $\vert \searrow \rangle$ for 1.
Since $\rho_+=\rho_\times=\rho$, if $\rho$
passes through the channel Bob will expect to receive $q\rho+(1-q)tr[\rho]\frac{1}{2}I$.
Thus if Alice attempts to cheat (i.e., entangle her secret system with Bob's qubit),
he will receive a separable state after applying the depolarizing channel ($\rho_{channel}=\sum_{i} p_i \rho_a^i\otimes \rho_b^i$).
Therefore the states of Alice will be disentangled from Bob's system, and more importantly he can determine if Alice has cheated or not.
To show this, two cases must be considered, Alice is honest and has not cheated, and Alice attempts to cheat.
For the first case, the probability of Bob receiving the same state as Alice sent is $q$.
Therefore the probability of Bob measuring the correct state is $\frac{q}{2}$ since the
probability of choosing the correct basis is $\frac{1}{2}$.
Bob should then expect to measure at least $\frac{q}{2}$ of the states correctly.
In the second case, Alice cheats and entangles her secret states with the committed qubits.
After Bob performs the depolarizing operation, the state will be disentangled as described previously,
and therefore Alice cannot change the state that Bob measures.
There is also no guarantee that the probability of Bob measuring the correct state is $\frac{q}{2}$,
so Alice may be exposed regarding the entanglements even if she does not change the committed bit.

A simple security analysis regarding our example using the Bennet and Brassard scheme is given below.
Alice prepares an entangled state $\vert a_0\rangle_A \vert 0\rangle_B+\vert a_1\rangle_A\vert 1\rangle_B$,
where the subscript $A$ means the state is controlled by Alice and $B$ controlled by Bob.
Alice then sends Bob's states to him, and when she wants to change her mind about the committed bit she performs a unitary transform followed by a
projective measurement on her own state (we can just assume a projective measurement).
The effect of the entanglement breaking channel on the system after Alice's measurement is
$\rho_b = \sum_{i} p_i \langle a_j \vert \rho_a^i \vert a_j\rangle \rho_b^i$, where $a_j$ is the basis for the projective measurement.
Thus in order for Alice to determine $\rho_b$ she needs to know the decomposition caused by the channel,
i.e., $\rho_{channel}=\sum_{i} p_i \rho_a^i\otimes \rho_b^i$, and therefore she needs to know the value of $q$ (which we assume is controlled by Bob).

Chailloux and Kerenidis \cite{20} provided lower bounds on an optimal quantum bit commitment (the bounds are tight and the upper bounds are close to the lower bounds), however they assumed that the operations in both the commitment and revealing phases are \textbf{unitary transforms} on Alice and Bob's quantum spaces. Here we take the same approach towards analysing the security of our system. For the Hiding property (i.e. the ability of Bob to guess the committed bit assuming a honest Alice), we know that without considering the channel Bob can guess the bit with probability $\frac{1}{2}+ \frac{\Delta(\sigma_0,\sigma_1)}{2}$ (where $\sigma_b$ is the density matrix assigned to 0 or 1). The effect of the channel, Bob's ability to cheat is then simply the maximum of his ability to distinguish the states with or without having them passed through the quantum channel, which is given by\\ $P_{Bcheat} = \frac{1}{2}+ Max(\frac{\Delta(\sigma_0,\sigma_1)}{2},\frac{\Delta(\mathbb{S}[\sigma_0],\mathbb{S}[\sigma_1])}{2})$ \\
Where $\mathbb{S}[.]$ is the effect of the channel. For Alice's cheating probability, consider the following.\\
We assume a cheating Alice prepares a state $\rho_{AB}$ and sends it to Bob so that just before Alice opens the bit, the state of that part of the system which Bob possesses is $\sigma_{B}=Tr_A(\alpha[I\otimes \mathbb{S}[\rho_{AB}]])$ where $\alpha[.]$ is Alice's operation on her own part of the system (i.e. A unitary transform followed by a measurement). Now assuming that $\mathbb{S}[.]$ is an entanglement breaking channel, we have $\sigma_{B}= \sum_{i} p_i Tr_A(A[\rho_a^i])\otimes \rho_b^i$. Assuming Alice wants Bob to measure 0, she should maximize the probability of Bob detecting $\sigma_0$, which is $F^2(\sigma_{Bend},\sigma_0)$, where $F$ is the fidelity. This means Alice must know the $p_i$ and properly choose the value of $Tr_A(A[\rho_a^i])$ (i.e. she also needs to know the $\rho_a^i$). This requires Alice to know the channel characteristics, but these are controlled by Bob and is kept secret by him.
\section*{Conclusions}

In this letter we have shown that by using an entanglement breaking channel,
the simple Bennett and Brassard bit commitment scheme can be made secure against EPR attacks.
We also presented an example of a depolarizing channel which is practically conceivable.
Only entanglement attacks were discussed, we leave the unconditional
security of these noise based systems as a topic for future research.

\pagebreak
\begin{figure}
\centering\includegraphics[scale=0.5]{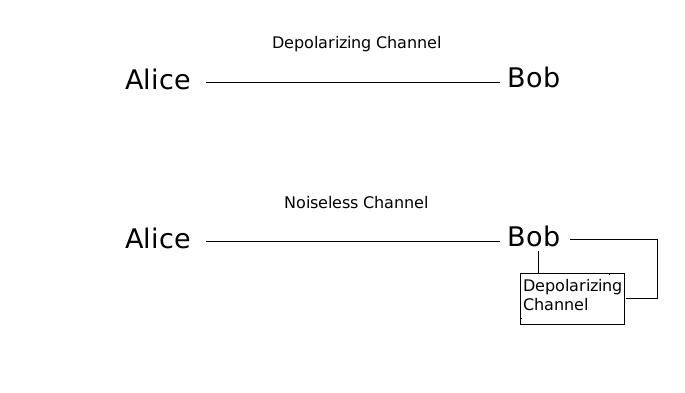}
\caption{Two possible implementations of the depolarizing channel for bit commitment.}
\end{figure}

\begin{thebibliography}{9}

\bibitem{1984-175-179}
Bennett, Charles H and Brassard, Gilles:
{Quantum cryptography: Public key distribution and coin tossing}, volume 11, Proceedings of IEEE International Conference on Computers Systems and Signal Processing, Bangalore, India, 1984, 175-179

\bibitem{PhysRevLett.78.3414}
{Unconditionally Secure Quantum Bit Commitment is Impossible},
{Apr},
{Phys. Rev. Lett.},
{10.1103/PhysRevLett.78.3414},
{Mayers, Dominic},
  {1997},
  {17},
{http://link.aps.org/doi/10.1103/PhysRevLett.78.3414},
{American Physical Society},
{3414--3417},
{78}

\bibitem{15}
 {Lenka Moravčíková and Mário Ziman},
 {Entanglement-annihilating and entanglement-breaking channels},
 {Journal of Physics A: Mathematical and Theoretical},
 {43},
 {27},
 {275306},
 {http://stacks.iop.org/1751-8121/43/i=27/a=275306},
 {2010}

\bibitem{16}
{Sergey N. Filippov, Tomas Rybar, Mario Ziman}
{Local two-qubit entanglement annihilating channels}
{arxiv.org/pdf/1110.3757}
\bibitem{17}
  {Quantum computation and quantum information},
  {2000},
  {0-521-63503-9},
  {Cambridge University Press},
{New York, NY, USA},


\bibitem{18}
{Evolution equation for quantum entanglement},
{Nat Phys},
{Vol. 4},
{No. 4.}
{(February 2008)},
{pp. 99-102.}

\bibitem{19}
 {Phys. Rev. A},
  {4},
  {Feb},
  {10.1103/PhysRevA.79.024303},
   {2},
 {Li, Zong-Guo and Fei, Shao-Ming and Wang, Z. D. and Liu, W. M.},
  {Evolution equation of entanglement for bipartite systems},
   {2009},
  {http://link.aps.org/doi/10.1103/PhysRevA.79.024303},
  {American Physical Society},
  {024303},
 {79}


\bibitem{Mayers99thetrouble},
{Dominic Mayers},
{The Trouble with Quantum Bit Commitment},
{Computing Research Repository (CoRR)},
{1999}

\bibitem{20},
{Andre Chailloux, Lordanis Kerenidis},
{Optimal bounds for quantum bit commitment},
{arxiv:1102.1678v1},
{2011}
\end{thebibliography}
\end{document}